# Nonlinear conformal-degree-preserving Dirac equations in 2+1 space-time

A. D. Alhaidari[a], U. Al Khawaja[a,b] and Y. H. Sabbah[b]

[a]Saudi Center for Theoretical Physics, P. O. Box 32741, Jeddah 21438, Saudi Arabia
[b]Physics Department, United Arab Emirates University, P. O. Box 15551, Al-Ain, United Arab Emirates

**Abstract**: We propose nonlinear Dirac equations where the conformal degree of the self-interaction terms are equal to that of the Dirac operator and the coupling parameters are dimensionless. As such, the massless equation is conformally invariant and preserves the conformal degree for both the linear and nonlinear components. In 2+1 space-time, we use these features to propose three- and four-parameter nonlinear Dirac equation models that might be useful for applications in 2D systems such as graphene sheets, ribbons and nanotubes.

## 1. Introduction:

In the language of conformally invariant theories, the conformal degree of a given field is the extent to (or order with) which it scales with the dimension of space and time. In $n+1$ Minkowski space-time, the kinetic energy term in the spinor Lagrangian is $i\bar{\psi}\not{\partial}\psi$, where $\not{\partial} = \sum_{\mu=0}^{\mu=n} \gamma^\mu \partial_\mu$ is the Dirac operator, $\{\gamma_\mu\}$ are the dimensionless Dirac gamma matrices, and $\bar{\psi} = \psi^\dagger \gamma_0$. Since the corresponding part of the action, which is the space-time integral $\int (i\bar{\psi}\not{\partial}\psi)d^{n+1}x$, is conformally invariant, then the conformal degree of $i\bar{\psi}\not{\partial}\psi$ is $-n-1$ making the conformal degree of the spinor field $-\frac{1}{2}n$. Recently, it was shown that a nonlinear conformally invariant massless Dirac equation in this space-time could be written, in the conventional relativistic units $\hbar = c = 1$, as [1,2]

$$i\not{\partial}\psi = \left[ \sum_k g_k \left(\bar{\psi}\Gamma_k\psi\right)^{1/n} \Lambda_k \right]\psi, \qquad (1)$$

where $\{\Gamma_k\}$ and $\{\Lambda_k\}$ are elements from the space of bilinear forms of the Dirac gamma matrices and $\{g_k\}$ are dimensionless coupling parameters. The sum of nonlinear terms in Eq. (1) runs over all independent coupling modes whose type and number depend on the dimensionality of space-time. These terms are obtained by stationary variations of the spinor field in the action. The nonlinear components of this action, which produce the right-hand side of Eq. (1), are constructed by requiring that their conformal degree be equal to $-n-1$ and by imposing relativistic invariance on the various types of coupling modes. These modes include scalar, vector, tensor, pseudo-scalar, etc. and could be written using objects such as the following

Scalar:     $S = \bar{\psi}\psi$,                                    (2a)
Vector:     $A_\mu = \bar{\psi}\gamma_\mu\psi$,                      (2b)

| Tensor: | $T_{\mu\nu} = \bar{\psi}\sigma_{\mu\nu}\psi$, | (2c) |
| Pseudo-Scalar: | $W = \bar{\psi}\gamma_5\psi$, | (2d) |
| Pseudo-Vector: | $P_\mu = \bar{\psi}\gamma_5\gamma_\mu\psi$, | (2e) |
| Pseudo-Tensor: | $Q_{\mu\nu} = \bar{\psi}\gamma_5\sigma_{\mu\nu}\psi$, | (2f) |

and so on, where $\gamma_5 = i\gamma_0\gamma_1...\gamma_n$, and $\sigma_{\mu\nu} = \frac{i}{4}[\gamma_\mu\gamma_\nu - \gamma_\nu\gamma_\mu]$. However, some of these couplings may become redundant in certain space-time dimensions. For example, in 2+1 dimension, $\sigma_{\mu\nu} \propto \varepsilon_{\mu\nu\lambda}\gamma^\lambda$ and $\gamma_5 \propto 1$, where $\varepsilon_{\mu\nu\lambda}$ is the completely antisymmetric tensor in 2+1 space-time. Therefore, the tensor, pseudo-scalar, pseudo-vector, and pseudo-tensor couplings are redundant. Now, the conformal degree of the spinor field in 2+1 space-time is $-1$. Thus, the independent nonlinear terms in the interaction Lagrangian with conformal degree $-3$ are the following

$$S^{3/2} = (\bar{\psi}\psi)^{3/2} \text{ and } (A^\mu A_\mu)^{3/4} = \left[(\bar{\psi}\gamma^\mu\psi)(\bar{\psi}\gamma_\mu\psi)\right]^{3/4}, \tag{3}$$

Consequently, the nonlinear *massive* Dirac equation (1) becomes

$$(i\slashed{\partial} - m)\psi = \left[\sum_{k=1}^{k=4} g_k\left(\sqrt{\bar{\psi}\Gamma_k\psi}\right)\Lambda_k\right]\psi, \tag{4}$$

where $m$ is the rest mass of the spinor particle. Now, for the *massless* case, this nonlinear Dirac equation is invariant under the action of the full conformal group [3]. However, if the fractional powers $3/2$ and $3/4$ in Eq. (3) are changed, which will destroy the overall conformal degree, then the symmetry of the massless equation is reduced and invariance is only under the smaller Weyl group [3].

## 2. The Models:

The minimum dimensional representation of the Dirac gamma matrices in 2+1 space-time is 2×2. These could be taken as: $\gamma_0 = \sigma_3 = \begin{pmatrix} 1 & 0 \\ 0 & -1 \end{pmatrix}$, $\gamma_1 = i\sigma_1 = i\begin{pmatrix} 0 & 1 \\ 1 & 0 \end{pmatrix}$, and $\gamma_2 = i\sigma_2 = \begin{pmatrix} 0 & 1 \\ -1 & 0 \end{pmatrix}$. Consequently, Eq. (4) becomes the following nonlinear matrix equation in 2+1 space-time with the most general coupling

$$i\partial_t \begin{pmatrix} \psi_+ \\ \psi_- \end{pmatrix} = \begin{pmatrix} m + \alpha_S \tilde{S} + \alpha_V \tilde{V} & \partial_x - i\partial_y + i\alpha_U \tilde{U} + \alpha_W \tilde{W} \\ -\partial_x - i\partial_y - i\alpha_U \tilde{U} + \alpha_W \tilde{W} & -m - \alpha_S \tilde{S} + \alpha_V \tilde{V} \end{pmatrix} \begin{pmatrix} \psi_+ \\ \psi_- \end{pmatrix}, \tag{5}$$

where $\{\alpha_S, \alpha_V, \alpha_U, \alpha_W\}$ are real dimensionless coupling parameters for the scalar coupling $S$ and the three-component vector coupling $\{A_0, A_1, A_2\} = \{V, U, W\}$. The self-interaction potential components are

$$\tilde{S}(t,x,y) = \sqrt{\bar{\psi}\psi} = \sqrt{|\psi_+|^2 - |\psi_-|^2}, \tag{6a}$$

$$\tilde{V}(t,x,y) = \sqrt{\bar{\psi}\gamma_0\psi} = \sqrt{|\psi_+|^2 + |\psi_-|^2}, \tag{6b}$$

$$\tilde{U}(t,x,y) = \sqrt{\bar{\psi}\gamma_1\psi} = \sqrt{i(\psi_+^*\psi_- - \psi_+\psi_-^*)}. \tag{6c}$$

$$\tilde{W}(t,x,y) = \sqrt{\bar{\psi}\gamma_2\psi} = \sqrt{\psi_+^*\psi_- + \psi_+\psi_-^*}. \tag{6d}$$

We may eliminate the $\tilde{U}$ component of the vector potential in the Dirac equation (5) by the nonlinear local gauge transformation $\psi \to e^{-i\alpha_U\theta}\psi$, where $\partial_x\theta = \tilde{U}$. Substituting these nonlinear terms into Eq. (5) and gauging away the $\tilde{U}$ term, we obtain the following three-parameter equation

$$i\partial_t \begin{pmatrix} \psi_+ \\ \psi_- \end{pmatrix} = \begin{pmatrix} m + \alpha_S \sqrt{|\psi_+|^2 - |\psi_-|^2} + \alpha_V \sqrt{|\psi_+|^2 + |\psi_-|^2} & \partial_x - i\partial_y + \alpha_W \sqrt{\psi_+^* \psi_- + \psi_+ \psi_-^*} \\ -\partial_x - i\partial_y + \alpha_W \sqrt{\psi_+^* \psi_- + \psi_+ \psi_-^*} & -m - \alpha_S \sqrt{|\psi_+|^2 - |\psi_-|^2} + \alpha_V \sqrt{|\psi_+|^2 + |\psi_-|^2} \end{pmatrix} \begin{pmatrix} \psi_+ \\ \psi_- \end{pmatrix} \quad (7)$$

We could have also made the problem more general by adding external linear potentials to the various coupling modes. However, we limit the present treatment to self-interaction. We could simplify Eq. (7) by proposing the following alternative four-parameter model that also preserves the conformal degree but violates Lorentz invariance

$$i\partial_t \begin{pmatrix} \psi_+ \\ \psi_- \end{pmatrix} = \begin{pmatrix} m + \alpha_+ |\psi_+| + \alpha_- |\psi_-| & \partial_x - i\partial_y + \beta_+ \psi_+ + \beta_- \psi_- \\ -\partial_x - i\partial_y + \beta_+ \psi_+^* + \beta_- \psi_-^* & -m + \alpha_+ |\psi_-| + \alpha_- |\psi_+| \end{pmatrix} \begin{pmatrix} \psi_+ \\ \psi_- \end{pmatrix}, \quad (8a)$$

and its "conjugate"

$$i\partial_t \begin{pmatrix} \psi_+ \\ \psi_- \end{pmatrix} = \begin{pmatrix} m + \alpha_+ |\psi_+| + \alpha_- |\psi_-| & \partial_x - i\partial_y + \beta_+ \psi_+^* + \beta_- \psi_-^* \\ -\partial_x - i\partial_y + \beta_+ \psi_+ + \beta_- \psi_- & -m + \alpha_+ |\psi_-| + \alpha_- |\psi_+| \end{pmatrix} \begin{pmatrix} \psi_+ \\ \psi_- \end{pmatrix}. \quad (8b)$$

where $\alpha_\pm$ and $\beta_\pm$ are four independent real dimensionless parameters. Violation of Lorentz invariance in nonlinear models might not be so dissolute. In fact, it was even suggested that at some fundamental level such violation could be tied to quantum non-linearity [4]. Moreover, we may think of these as effective models that describe a given physical phenomenon rather than a fundamental interaction. On the other hand, the following three-parameter model is Lorentz invariant but does not preserve the conformal degree in 2+1 space-time [1]

$$i\partial_t \begin{pmatrix} \psi_+ \\ \psi_- \end{pmatrix} = \begin{pmatrix} m + \alpha_+ |\psi_+|^2 + \alpha_- |\psi_-|^2 & \partial_x - i\partial_y + \alpha_W \left( \psi_+^* \psi_- + \psi_+ \psi_-^* \right) \\ -\partial_x - i\partial_y + \alpha_W \left( \psi_+^* \psi_- + \psi_+ \psi_-^* \right) & -m + \alpha_+ |\psi_-|^2 + \alpha_- |\psi_+|^2 \end{pmatrix} \begin{pmatrix} \psi_+ \\ \psi_- \end{pmatrix}, \quad (9)$$

where the coupling parameters have length dimensions (i.e., their conformal degree is +1). This could have a negative impact on the renormalizability of the theory.

The Dirac equation (5) is written in Cartesian coordinates, which is suitable for the treatment of problems with rectangular configuration. On the other hand, problems with circular configuration are described properly in cylindrical coordinates $(r,\theta)$. To simplify the writing of such an equation, we may follow a procedure described in Ref. [5] whereby a local 2×2 transformation $\Lambda(r,\theta)$ is applied to the Dirac equation such that the cylindrical projection of the gamma matrices $\sigma_r = \vec{\sigma} \cdot \hat{r}$ and $\sigma_\theta = \vec{\sigma} \cdot \hat{\theta}$ are mapped into their canonical Cartesian representations $\sigma_1$ and $\sigma_2$, respectively. That is, $\Lambda(\sigma_r)\Lambda^{-1} = \sigma_1$ and $\Lambda(\sigma_\theta)\Lambda^{-1} = \sigma_2$. Moreover, we require that the transformed Dirac Hamiltonian matrix be Hermitian. One may refer to Ref. [5] for the details of the procedure where one obtains $\Lambda(r,\theta) = \sqrt{r} e^{\frac{i}{2}\theta\sigma_3}$ and the nonlinear Dirac equation (7) takes the following form in cylindrical coordinates

$$i\partial_t \begin{pmatrix} \phi_+ \\ \phi_- \end{pmatrix} = \begin{pmatrix} m + \frac{\alpha_S}{\sqrt{r}} \sqrt{|\phi_+|^2 - |\phi_-|^2} + \frac{\alpha_V}{\sqrt{r}} \sqrt{|\phi_+|^2 + |\phi_-|^2} & \partial_r - \frac{i}{r}\partial_\theta + \frac{\alpha_W}{\sqrt{r}} \sqrt{\phi_+^* \phi_- + \phi_+ \phi_-^*} \\ -\partial_r - \frac{i}{r}\partial_\theta + \frac{\alpha_W}{\sqrt{r}} \sqrt{\phi_+^* \phi_- + \phi_+ \phi_-^*} & -m - \frac{\alpha_S}{\sqrt{r}} \sqrt{|\phi_+|^2 - |\phi_-|^2} + \frac{\alpha_V}{\sqrt{r}} \sqrt{|\phi_+|^2 + |\phi_-|^2} \end{pmatrix} \begin{pmatrix} \phi_+ \\ \phi_- \end{pmatrix} \quad (10)$$

where $\phi = \Lambda \psi$, which has a conformal degree of $-\frac{1}{2}$. Therefore, the two-component spinor wavefunction becomes $\psi(t,r,\theta) = \frac{1}{\sqrt{r}} e^{-\frac{i}{2}\theta\sigma_3} \phi(t,r,\theta)$. That is, $\phi_\pm = \sqrt{r} e^{\pm\frac{i}{2}\theta} \psi_\pm$ and the density of state $\rho(t) = \int (\psi^\dagger \psi) r \, dr \, d\theta = \int (\phi^\dagger \phi) \, dr \, d\theta$. Additionally, the model given by Eq. (8a) and its conjugate in Eq. (8b) become

$$i\partial_t \begin{pmatrix} \phi_+ \\ \phi_- \end{pmatrix} = \begin{pmatrix} m + \frac{\alpha_+}{\sqrt{r}}|\phi_+| + \frac{\alpha_-}{\sqrt{r}}|\phi_-| & \partial_r - \frac{i}{r}\partial_\theta + \frac{1}{\sqrt{r}}(\beta_+ \phi_+ + \beta_- \phi_-) \\ -\partial_r - \frac{i}{r}\partial_\theta + \frac{1}{\sqrt{r}}(\beta_+ \phi_+^* + \beta_- \phi_-^*) & -m + \frac{\alpha_+}{\sqrt{r}}|\phi_-| + \frac{\alpha_-}{\sqrt{r}}|\phi_+| \end{pmatrix} \begin{pmatrix} \phi_+ \\ \phi_- \end{pmatrix}. \quad (11a)$$

$$i\partial_t \begin{pmatrix} \phi_+ \\ \phi_- \end{pmatrix} = \begin{pmatrix} m + \frac{\alpha_+}{\sqrt{r}}|\phi_+| + \frac{\alpha_-}{\sqrt{r}}|\phi_-| & \partial_r - \frac{i}{r}\partial_\theta + \frac{1}{\sqrt{r}}(\beta_+ \phi_+^* + \beta_- \phi_-^*) \\ -\partial_r - \frac{i}{r}\partial_\theta + \frac{1}{\sqrt{r}}(\beta_+ \phi_+ + \beta_- \phi_-) & -m + \frac{\alpha_+}{\sqrt{r}}|\phi_-| + \frac{\alpha_-}{\sqrt{r}}|\phi_+| \end{pmatrix} \begin{pmatrix} \phi_+ \\ \phi_- \end{pmatrix}. \quad (11b)$$

To be able to obtain stationary solutions of the form $\psi_\pm(t,x,y) = e^{-i\varepsilon t} \chi_\pm(x,y)$, we need to make sure that with this substitution time becomes separable in the above models. It is evident that the three-parameter Lorentz invariant model of Eq. (7) and its cylindrical version of Eq. (10) are compatible with this requirement. On the other hand, the four-parameter Lorentz-symmetry-violating model given by Eqs. (8) or Eqs. (11) are not. Consequently, we modify this latter model to read as follows in Cartesian coordinates

$$i\partial_t \begin{pmatrix} \psi_+ \\ \psi_- \end{pmatrix} = \begin{pmatrix} m + \alpha_+|\psi_+| + \alpha_-|\psi_-| & \partial_x - i\partial_y + \beta_+|\psi_+| + \beta_-|\psi_-| \\ -\partial_x - i\partial_y + \beta_+|\psi_+| + \beta_-|\psi_-| & -m + \alpha_+|\psi_-| + \alpha_-|\psi_+| \end{pmatrix} \begin{pmatrix} \psi_+ \\ \psi_- \end{pmatrix}. \quad (12)$$

Whereas in cylindrical coordinates, this same model reads as follows

$$i\partial_t \begin{pmatrix} \phi_+ \\ \phi_- \end{pmatrix} = \begin{pmatrix} m + \frac{\alpha_+}{\sqrt{r}}|\phi_+| + \frac{\alpha_-}{\sqrt{r}}|\phi_-| & \partial_r - \frac{i}{r}\partial_\theta + \frac{1}{\sqrt{r}}(\beta_+|\phi_+| + \beta_-|\phi_-|) \\ -\partial_r - \frac{i}{r}\partial_\theta + \frac{1}{\sqrt{r}}(\beta_+|\phi_+| + \beta_-|\phi_-|) & -m + \frac{\alpha_+}{\sqrt{r}}|\phi_-| + \frac{\alpha_-}{\sqrt{r}}|\phi_+| \end{pmatrix} \begin{pmatrix} \phi_+ \\ \phi_- \end{pmatrix}. \quad (13)$$

## 3. Sample Solutions:

As illustration of the above findings, we consider two relevant physical configurations in which one of the space coordinates become cyclic (i.e., the interaction is independent of that coordinate):

1. In Cartesian coordinates, we consider $y$-independent interactions. That is, $y$ becomes the cyclic coordinate and the two spinor components could be written as $\psi_\pm(t,x,y) = e^{i(ky-\varepsilon t)} \chi_\pm(x)$, where the real parameter $k$ is the wavenumber along the $y$-axis. Such solutions are relevant to physical configurations such as nano-ribbons and nano-strips.
2. In cylindrical coordinates, we consider cylindrically symmetric interactions. That is, $\theta$ becomes the cyclic coordinate and the two-component spinor wave function becomes

$$\psi(t,r,\theta) = \frac{1}{\sqrt{r}} e^{\frac{i}{2}\theta\sigma_3} e^{i\kappa\theta} \phi(t,r) = \frac{1}{\sqrt{r}} e^{\frac{i}{2}\theta\sigma_3} e^{i(\kappa\theta-\varepsilon t)} \chi(r), \quad (14)$$

where $\kappa$ is the azimuthal wave number. These solutions are relevant to nano-sheets in the form of disks and sectors. For nano-disks and nanotubes, where $0 \leq \theta < 2\pi$, continuity dictates that $\kappa$ becomes the azimuthal quantum number $\kappa = \pm\frac{1}{2}, \pm\frac{3}{2}, \pm\frac{5}{2},\ldots$.

In Table 1, we list exact analytic solutions in cylindrical and Cartesian coordinates for special values of the model parameters. The importance of these solutions is not only in showing that the model can be solved analytically and exactly, but also in the diversity of physical configurations of the model that differ significantly according to the choice of values of those parameters. Therefore, in principle, there are many other analytic solutions beyond those obtained here.

## 4. Conclusion:

In conclusion, we imposed the constraint that all components of the massless Dirac equation (linear and nonlinear) in 2+1 space-time have the same conformal degree of –1 and that the coupling parameters are dimensionless. We showed that these constraints together with Lorentz invariance lead to the nonlinear three-parameter Dirac equation model given by Eq. (7) where only the mass term violates conformal invariance. We simplified the model by breaking Lorentz invariance, which resulted in the four-parameter model (and its conjugate) shown in Eq. (8a) and Eq. (8b). Additionally, to make this latter model separable in time so that stationary solutions can be obtained, we modified it as shown in Eq. (12) and Eq. (13). Sample analytic solutions were obtained. The important issues of integrability, exact solvability and range of stability of solutions of these models are left for future studies.

## References

[1] A. D. Alhaidari, "*Nonlinear spin and pseudo-spin symmetric Dirac equations*", Int. J. Theor. Phys. **53**, 685 (2014)
[2] U. Al Khawaja, "*Exact localized and oscillatory solutions of the nonlinear spin and pseudospin symmetric Dirac equations*", Phys. Rev. A **90**, 052105 (2014)
[3] W. I. Fushchich and W. M. Shtelen, "*On some exact solutions of the nonlinear Dirac equation*", J. Phys. A **16**, 271 (1983)
[4] R. Parwani, "*An information-theoretic link between space-time symmetries and quantum linearity*", Ann. Phys. **315**, 419 (2005)
[5] A. Jellal, A. D. Alhaidari, and H. Bahlouli, "*Confined Dirac fermions in a constant magnetic field*", Phys. Rev. A **80**, 012109 (2009) pp. 2-3

## Table Caption:

**Table 1**: Sample analytic solutions in cylindrical and Cartesian coordinates for the model equations shown in the first column and for various special values of the model parameters.

**Table 1**

| Model | Wavefunction Components | Parameters | Amplitude Functions |
|---|---|---|---|
| Eq. (13) | $\psi_\mp = f_\mp(r)e^{i(\varepsilon t + \gamma_\mp(\theta))}$ | $m = \varepsilon = 0$<br>$\alpha_\pm = 0$<br>$\gamma_\pm(\theta) = 0$ | $f_+(r) = \dfrac{c\sqrt{\beta_-}\tan(2c\sqrt{\beta_+\beta_-}\,r)}{\sqrt{\beta_+}}$<br>$f_-(r) = \dfrac{c\sqrt{\beta_+}\cot(2c\sqrt{\beta_+\beta_-}\,r)}{\sqrt{\beta_-}}$ |
| | | $m = \varepsilon = 0$<br>$\alpha_+ = 0$<br>$\beta_\pm = 0$<br>$\gamma_\pm(\theta) = 0$ | $f_+(r) = \dfrac{c_1 c_2 e^{2c_1\alpha_-\sqrt{r}}}{1 + c_2 e^{2c_1\alpha_-\sqrt{r}}}$<br>$f_-(r) = \dfrac{c_1}{1 + c_2 e^{2c_1\alpha_-\sqrt{r}}}$ |
| Eq. (7) | $\psi_\pm = f_\pm(x)e^{i(ky-\varepsilon t)}$ | $\alpha_S = 0$<br>$\alpha_V = 0$<br>$\varepsilon = m$<br>$k = 0$ | $f_+(x) = -\dfrac{3}{2\sqrt{2m}\,\alpha_W x^{\frac{3}{2}}}$<br>$f_-(x) = -\dfrac{3\sqrt{m}}{2\sqrt{2}\,\alpha_W x^{\frac{1}{2}}}$ |
| Eq. (12) | $\psi_\mp = f_\mp(x)e^{i(ky-\varepsilon t)}$ | $\varepsilon = m$<br>$k = 0$<br>$\beta_\pm = 0$<br>$\alpha_- = 0$ | $f_+(x) = \dfrac{3m^2 x}{\alpha_+(1 + m^3 x^3)}$<br>$f_-(x) = \dfrac{3m}{\alpha_+(1 + m^3 x^3)}$ |